\documentclass[letterpaper]{article} % DO NOT CHANGE THIS
\usepackage{aaai2026}  % DO NOT CHANGE THIS
\usepackage{times}  % DO NOT CHANGE THIS
\usepackage{helvet}  % DO NOT CHANGE THIS
\usepackage{courier}  % DO NOT CHANGE THIS
\usepackage[hyphens]{url}  % DO NOT CHANGE THIS
\usepackage{graphicx} % DO NOT CHANGE THIS
\usepackage{adjustbox}
\usepackage{natbib}  % DO NOT CHANGE THIS AND DO NOT ADD ANY OPTIONS TO IT
\usepackage{caption} % DO NOT CHANGE THIS AND DO NOT ADD ANY OPTIONS TO IT
\usepackage{algorithm}
\usepackage{algorithmic}

\usepackage{newfloat}
\usepackage{listings}

\DeclareCaptionStyle{ruled}{labelfont=normalfont,labelsep=colon,strut=off} % DO NOT CHANGE THIS
\floatstyle{ruled}
\newfloat{listing}{tb}{lst}{}
\floatname{listing}{Listing}
\newcommand{\model}{\texttt{JAM}}
\newcommand{\dataset}{\texttt{JAME}}

\newcommand{\specialtoken}[1]{\texttt{\textsc{#1}}}

\title{\model{}: A Tiny Flow-based Song Generator with Fine-grained Controllability and Aesthetic Alignment}
\author{
    Renhang Liu\textsuperscript{\rm 1},
    Chia-Yu Hung\textsuperscript{\rm 1},
    Navonil Majumder\textsuperscript{\rm 1},
    Taylor Gautreaux\textsuperscript{\rm 2},
    Amir Ali Bagherzadeh\textsuperscript{\rm 2},
    Chuan Li\textsuperscript{\rm 2},
    Dorien Herremans\textsuperscript{\rm 1},
    Soujanya Poria\textsuperscript{\rm 1}
}
\affiliations{
    \textsuperscript{\rm 1}Singapore University of Technology and Design\\
    \{renhang\_liu, chiayu\_hung, navonil\_majumder, dorien\_herremans, sporia\}@sutd.edu.sg\\
    \textsuperscript{\rm 2}Lambda Labs\\
    \{taylor.gautreaux,     amirali.zadeh, c\}@lambda.ai
}

\usepackage{graphicx}
\usepackage{caption}
\usepackage{algorithm}
\usepackage{algorithmic}
\usepackage{multirow}
\usepackage{amsfonts,amsmath,amssymb}
\usepackage{subcaption}
\usepackage[capitalize]{cleveref}
\usepackage{enumitem}
\usepackage{pifont}
\usepackage{newfloat}
\usepackage{listings}
\usepackage{soul}
\usepackage[table,xcdraw]{xcolor}
\usepackage{tikz}
\usepackage{arydshln}
\usepackage{tabularx}
\usepackage{booktabs}
\usepackage{tcolorbox}
\usepackage{mdframed}
\usepackage{fancyvrb}
\usepackage{adjustbox}
\definecolor{lightyellow}{HTML}{ffe599}
\definecolor{green}{HTML}{34a853}
\definecolor{lightcornflowerblue}{HTML}{c9daf8}
\definecolor{darkyellow}{rgb}{0.85, 0.65, 0.13}

\definecolor{nmcolor}{RGB}{255, 25, 26}

\let\realcite\cite
\renewcommand{\cite}[1]{\ifx.#1.\hl{[?]}\else\realcite{#1}\fi}
\definecolor{jamcolor}{HTML}{005F73} % Teal-ish color
\definecolor{linkbg}{HTML}{F1FAEE}   % Soft background
\definecolor{lightergray}{RGB}{217, 219, 221}
\definecolor{psy}{RGB}{250, 230, 254}
\newcommand{\mystack}[2]{\begin{tabular}{@{}l@{}} #1 \\ ~~#2 \end{tabular}}
\newcommand{\mytstack}[3]{\renewcommand{\arraystretch}{0.7}\begin{tabular}{@{}l@{}} #1 \\ \tiny #2 #3 \end{tabular}}

\makeatletter
\AtBeginDocument{
\let\@oldmaketitle\@maketitle
\renewcommand{\@maketitle}{\@oldmaketitle
  \vspace{-2pt}
  \begin{center}
   \includegraphics[width=\linewidth]{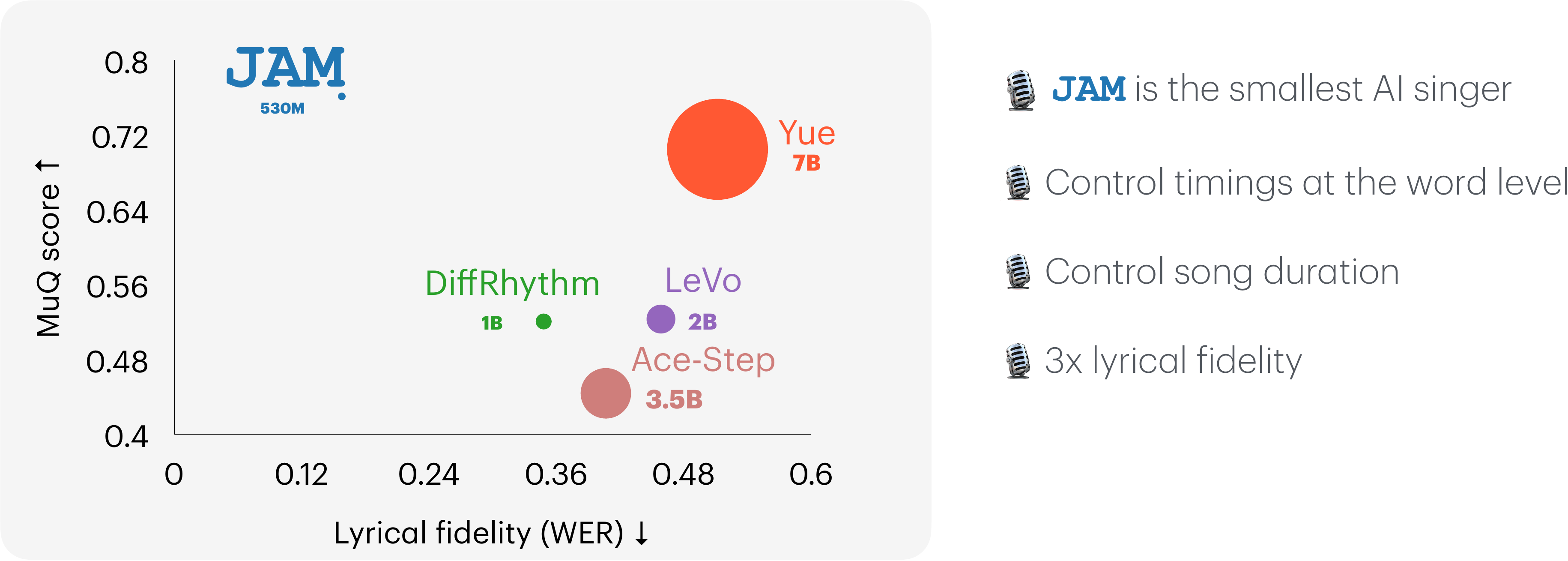}
\textbf{\textsf{Project \textsc{Jamify}}}: 
\texttt{\textcolor{jamcolor}{\url{https://declare-lab.github.io/jamify}}} \\[2pt]
\textbf{\textsf{Model}}: 
\texttt{\textcolor{jamcolor}{\url{https://huggingface.co/declare-lab/JAM-0.5}}} \\[2pt]
\textbf{\textsf{Github}}: 
\texttt{\textcolor{jamcolor}{\url{https://github.com/declare-lab/jamify}}}
\\[3pt]
{\color{red} {\textbf{\textsf{Disclaimer}}: \model{} is intended for use by music professionals, as it requires desired total and word-level duration inputs that a music expert is able to provide. Use by non-experts or without accurate word timings may result in suboptimal outputs, including vocal-accompaniment mismatches and artifacts. These issues can be mitigated with a duration predictor.}}
  \end{center}  
 }
}
\makeatother
\begin{document}

\maketitle
\begin{tikzpicture}[remember picture,overlay,shift={(current page.north west)}]
\node[anchor=north west,xshift=0.4cm,yshift=-1.9cm]{\scalebox{1}[1]{\includegraphics[width=1.5cm]{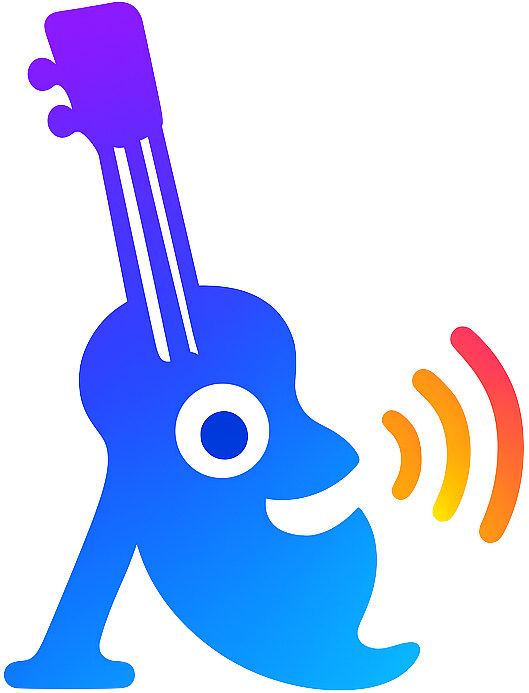}}};
%\vspace{-4em}
\end{tikzpicture}
\begin{abstract}
Diffusion and flow-matching models have revolutionized automatic text-to-audio generation in recent times. These models are increasingly capable of generating high quality and faithful audio outputs capturing to speech and acoustic events. However, there is still much room for improvement in creative audio generation that primarily involves music and songs. Recent open lyrics-to-song models, such as, DiffRhythm, ACE-Step, and LeVo, have set an acceptable standard in automatic song generation for recreational use. However, these models lack fine-grained word-level controllability often desired by musicians in their workflows. To the best of our knowledge, our flow-matching-based \model{} is the first effort toward endowing word-level timing and duration control in song generation, allowing fine-grained vocal control. To enhance the quality of generated songs to better align with human preferences, we implement aesthetic alignment through Direct Preference Optimization, which iteratively refines the model using a synthetic dataset, eliminating the need for manual data annotations. Furthermore, we aim to standardize the evaluation of such lyrics-to-song models through our public evaluation dataset \dataset{}. We show that \model{} outperforms the existing models in terms of the music-specific attributes.
\end{abstract}
\section{Introduction}

Music plays an essential role in human culture: it brings people together, expresses emotions, embodies cultural elements, and through them enriches our daily lives via shared experiences. From ancient rituals to modern celebrations, music has shaped our social bonds and personal well–being \cite{freeman1998neurobiological,agres2021music}.

Creating music, however, is a complex and time–consuming process that often demands extensive effort from skilled musicians. Recent advances in neural generative models have made AI–generated music a reality, providing composers with reliable first drafts that accelerate the creative workflow.

The recent audio-based generative music AI models can be grouped into three main trends:
\begin{itemize}
  \item \emph{Singing voice generation} focuses on producing expressive vocal performances from given lyrics and musical notes—often using techniques like voice cloning—without generating instrumental accompaniment.
  \item \emph{Textual description to music generation} learns to translate prompts that may include instructions about mood, desired musical features, or instrumentation into fully synthesized music, but typically does not handle vocals. These works are very similar to text-to-audio generative models, such as, Tango series models~\cite{tango,tango2,tangoflux}, AudioLDM~\cite{liu2024audioldm2learningholistic}, and Stable Audio Open~\cite{evans2024stableaudioopen}. These models are capable of generating both sound effects and instrumental music from textual prompts.
  \item \emph{Lyrics-to-song generation} combines both singing voice and accompaniment to produce full songs, ensuring semantic coherence between lyrics and music as well as acoustic harmony across vocals and instruments.
\end{itemize}
Among these, lyrics-to-song generation presents unique challenges in aligning linguistic content with musical structure, preserving prosody, and generating high–quality audio across longer durations.

Existing approaches to lyrics-to-song generation fall into two broad categories. \emph{Autoregressive models} generate audio tokens sequentially, which allows them to maintain strong long–range coherence and to incorporate explicit musical style controls. However, their step–by–step decoding can be prohibitively slow for practical use. In contrast, \emph{diffusion–based methods} iteratively denoise latent representations, offering high audio fidelity, flexible conditioning on melody, rhythm, and timbre, and natural support for editing and style transfer. Despite these advantages, diffusion models can still struggle with generation speed and fine–grained alignment between lyrics and audio.

Recent systems such as DiffRhythm~\cite{ning2025diffrhythmblazinglyfastembarrassingly}, YuE~\cite{yuan2025yuescalingopenfoundation}, LeVo~\cite{lei2025levohighqualitysonggeneration}, and ACE-Step~\cite{gong2025acestepstepmusicgeneration} have demonstrated impressive results, but they share several limitations:
\begin{enumerate}
  \item \textbf{Large model size:} All are based on hundreds of millions to billions of parameters, leading to slow inference regardless of the generation paradigm. This also makes the model more resource hungry.
  \item \textbf{Coarse timing control:} While DiffRhythm allows specifying sentence–level start times, none support word– or phoneme–level alignment, limiting the user’s ability to shape prosody and rhythm precisely.
  \item \textbf{Weak lyric fidelity:} High word error rates (WER) and phoneme error rates (PER) indicate that the models often misalign or omit lyric content.
  \item \textbf{Lack of duration control:} Without explicit control over the overall song duration and inter–word pauses, users cannot easily shape the global structure or pacing of the generated song.
\end{enumerate}

To address these challenges, we introduce \model{}, a rectified–flow based model for lyrics to song generation. \model{} is a 530M–parameter conditional flow–matching model~\cite{lipman2023flowmatchinggenerativemodeling} built on 16 LLaMA–style Transformer layers as the Diffusion Transformer (DiT) backbone~\cite{peebles2023scalablediffusionmodelstransformers}. It is jointly conditioned on:
\begin{itemize}
  \item \emph{Lyrics}, with fine–grained word– and phoneme–level start/end times for precise prosody control.
  \item \emph{Target duration}, guiding the model on the overall song length and the spacing between vocal phrases.
  \item \emph{Style prompt}, which can be either a reference audio clip or a text description, to capture desired timbral and structural characteristics.
\end{itemize}
\model{} generates full songs at 44.1kHz for up to 3 minutes and 50 seconds by learning a rectified flow trajectory through the latent space of a variational autoencoder (VAE).

Our contributions are:
\begin{enumerate}
  \item \textbf{Compact architecture:} At 530M parameters, \model{} is less than half the size of the next smallest system (Diffrhythm-1.1B), enabling faster inference and is less resource demanding.
  \item \textbf{Fine–grained alignment:} By accepting word– and phoneme–level timing inputs, \model{} lets users control the exact placement of each vocal sound, improving rhythmic flexibility and expressive timing.
  \item \textbf{Enhanced lyric fidelity:} This precise alignment reduces WER and PER by over 3× compared to prior work, as the model can directly attend to phoneme boundaries and correct misalignments.
  \item \textbf{Global duration control:} Our novel duration mechanism not only sets the inter–word pacing implicitly (it is controlled explicitly through lyric timing) but also specifies how much instrumental introduction and coda to generate, giving composers full control over song structure.
\item \textbf{Rigorous evaluation:} Assessing prior methods is difficult when their training data is undisclosed. To avoid data contamination, we compiled lyrics for 250 tracks released after the models’ training cut‑offs—ensuring neither \model{} nor any baseline had access to them. Additionally, these tracks span a variety of genres, allowing us to evaluate performance across different musical styles.
\item \textbf{Aesthetic alignment:} Most prior systems (with the exception of LeVo) lack any mechanism for aligning model outputs to human aesthetic preferences. In text–to–audio work such as Tango2 and Tangoflux, preference alignment has proven effective, and LeVo recently adapted this idea for song generation. However, LeVo relies on a manually annotated preference dataset, which incurs significant human effort. Inspired by Tango2 and Tangoflux, we instead use automated song–quality models like SongEval to generate synthetic preference labels. We further apply this alignment in multiple rounds, yielding additional performance gains.
\end{enumerate}

With these advances, \model{} offers an efficient, controllable, and high–fidelity solution for turning lyrics into complete songs, paving the way for AI–assisted composition in both professional and amateur settings.

\section{Related Works}
\subsection{Music Generation}
Music generation aims to generate coherent and aesthetically pleasing musical content, either as raw audio, conditioned on various inputs such as text, lyrics, and style. Several recent works~\cite{evans2024stableaudioopen, lam2023efficientneuralmusicgeneration,chen2023musicldmenhancingnoveltytexttomusic,liu2024audioldm2learningholistic} adopt diffusion models~ \cite{ho2020denoisingdiffusionprobabilisticmodels,song2022denoisingdiffusionimplicitmodels} to generate music based on conditioning.  Mustango~\cite{melechovsky2024mustangocontrollabletexttomusicgeneration} and MusicLDM~\cite{chen2023musicldmenhancingnoveltytexttomusic} include musical information in text prompts such as chords, beats, and tempo to provide additional control over the musical structure and style. Other prominent approaches~\cite{copet2024simplecontrollablemusicgeneration,agostinelli2023musiclmgeneratingmusictext} employ an autoregressive transformer for generation, which encodes music as discrete token sequences by the Vector Quantised Variational Autoencoder (VQ-VAE)~\cite{oord2018neuraldiscreterepresentationlearning}. However, such approaches tend to have a high computational cost for generating longer duration music, making it unsuitable for long length music generation.

\subsection{Song Generation}
Song generation aims to produce realistic vocals accompanied by background music. Recent approaches adopt a two-stage process, which first generates vocals, followed by accompaniment music, where some approaches generate both vocals and accompaniment simultaneously~\cite{yuan2025yuescalingopenfoundation, lei2025levohighqualitysonggeneration, ning2025diffrhythmblazinglyfastembarrassingly,hong2024texttosongcontrollablemusicgeneration} have adopted either autoregressive or diffusion-based frameworks for this task. SongCreator~\cite{lei2024songcreatorlyricsbaseduniversalsong} introduces a dual-stream token generation approach, separately modeling music and vocal streams to enhance the overall musicality of the generated song. SongGen~ \cite{liu2025songgensinglestageautoregressive}, YuE~ \cite{yuan2025yuescalingopenfoundation}, and LeVo~ \cite{lei2025levohighqualitysonggeneration} generate both vocal and accompaniment tokens, with SongGen employing a codebook-delay mechanism inspired by MusicGen. DiffRhythm and DiffRhythm+~\cite{chen2025diffrhythmcontrollableflexiblefulllength,ning2025diffrhythmblazinglyfastembarrassingly} leverage diffusion models for song generation conditioned on both style and lyrics; the latter further incorporates preference optimization techniques originally proposed in Tango2~\cite{majumder2024tango2aligningdiffusionbased} and TangoFlux~\cite{hung2025tangofluxsuperfastfaithful}.

% \section{Preliminaries}

\section{Method}
\begin{figure*}
    \centering
    \includegraphics[width=\textwidth]{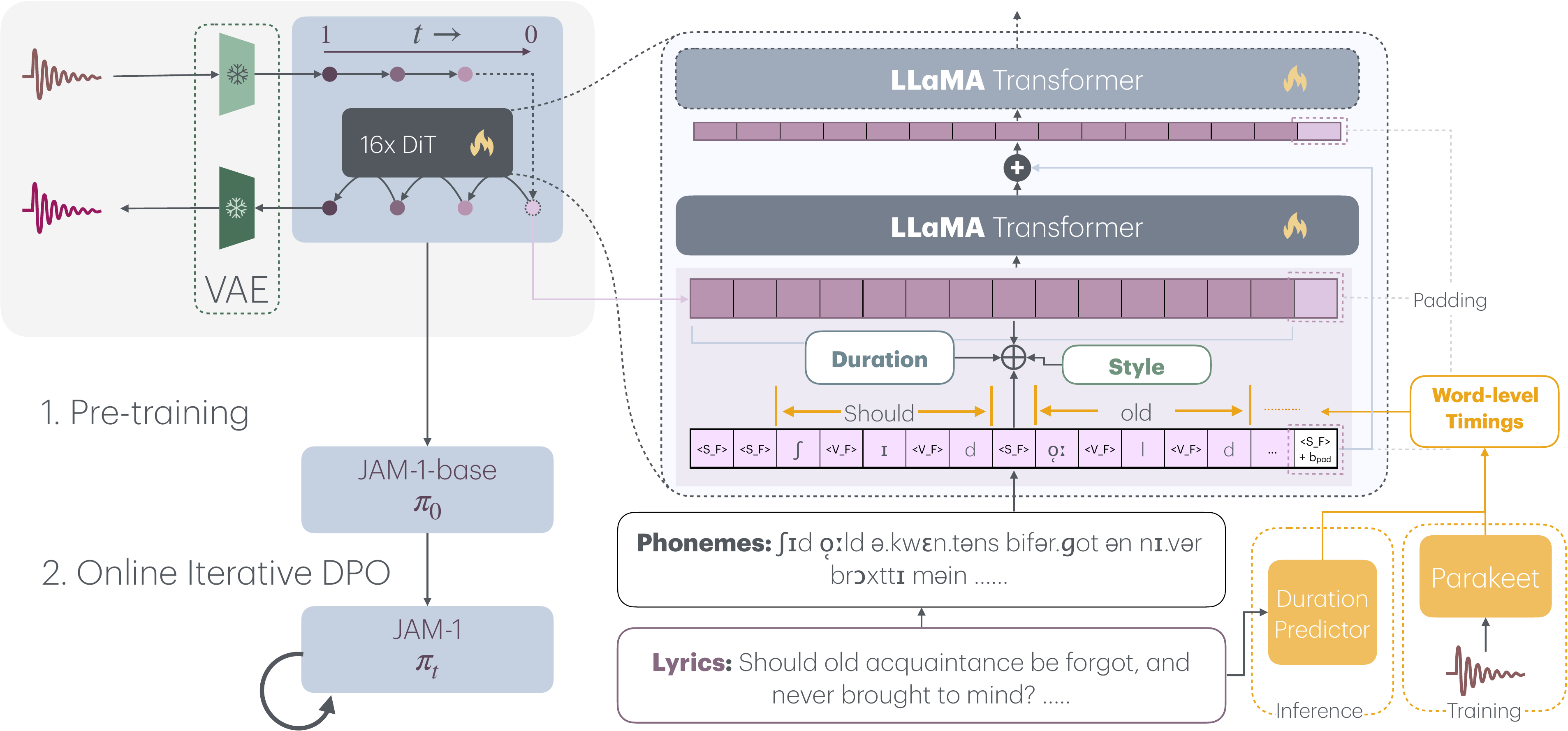}
    \caption{A depiction of our proposed architecture and training pipeline.}
    \label{fig:model}
\end{figure*}
\subsection{Model Architecture Overview}
\model{} is a $530$M parameter conditional flow-matching model~\cite{lipman2023flowmatchinggenerativemodeling} consisting of 16 LLaMA-like transformer layers as the Diffusion Transformer (DiT) \cite{peebles2023scalablediffusionmodelstransformers} backbone, conditioned on lyrics, target duration, and style prompt which can either be a clip of reference audio or a text description to generate full songs at 44.1kHz up to 3 minutes and 50 seconds. \model{} learns a rectified flow trajectory to a latent representation encoded by a variational autoencoder (VAE). Our key contributions include word-level timing controllability, duration modeling, and iterative automated aesthetic alignment.

\subsection{Training Pipeline}
Our training pipeline is composed of three stages:
\begin{enumerate}
    \item \textbf{Pre-training}: Train the model to generate 90-second song clips with randomly cropped clips from the training dataset.
    \item \textbf{Fine-tuning}: Fine-tune the pre-trained model for full-song generation with full-length songs.
    \item \textbf{Preference Alignment}: Post-train the full-song generator using iterative direct preference optimization (DPO) \cite{rafailov2024directpreferenceoptimizationlanguage} iteratively with candidates selected by averaged SongEval \cite{yao2025songevalbenchmarkdatasetsong} scores across different criteria. We described the detailed process in \Cref{sec:alignment}.
\end{enumerate}

\subsection{Audio Encoding}
\model{} uses the VAE encoder from Stable Audio Open~\cite{evans2024stableaudioopen} combined with the VAE decoder from DiffRhythm \cite{ning2025diffrhythmblazinglyfastembarrassingly}. The DiffRhythm decoder, initialized with Stable Audio Open's decoder's weights, is trained to enable the reconstruction of lossless-quality audio directly from latent representations of MP3-compressed (lossy) audio, while remaining fully compatible with the Stable Audio Open encoder's latent space. The encoder is capable of encoding $44.1$kHz stereo audio waveforms into a compressed latent representation. Given the stereo audio \(x \in \mathbb{R}^{2\times d \times sr}\), it encodes $x$ into a latent representation $z \in \mathbb{R}^{l \times c}$, where $l=f * d$ is the latent sequence length and $f$ is the latent frame rate, $c$ is the channel size. The decoder transforms the latent representation $z$ back to the original stereo waveform $x$. Both the VAE encoder and decoder are kept frozen during all training stages of \model{{}.

\subsection{Flow Matching}
Flow Matching (FM)~\cite{lipman2023flowmatchinggenerativemodeling} offers a simulation-free framework for training continuous normalizing flows (CNFs). In contrast to score-based diffusion models, which learn the gradient of the log-density $\nabla \log p_t(x)$, FM directly models a time-dependent vector field $v_t(x)$ that transports samples along trajectories from a simple prior distribution $p_0$ to a complex target distribution $p_1$, or vice versa. Training is typically formulated as a regression task, minimizing the difference between the predicted vector field and a reference vector field constructed from samples of both $p_0$ and $p_1$. This approach leads to a simpler objective function, potentially faster convergence, and greater training stability compared to score-based methods. Once trained, sampling from FM models is performed by numerically integrating an ordinary differential equation (ODE) defined by the learned vector field.

\subsubsection{Training.}
Given a latent representation of a song sample $z_1 \in \mathbb{R}^{l \times c}$, a noise sample \(z_0 \sim \mathcal{N}(0, 1)^{l \times c}\), time-step  $t \in [0,1]$, we can construct a training sample \(z_t\) where the model learns to predict a velocity $v_t = \frac{dz_t}{dt}$ that guides \(z_t\) to \(z_1\). During \model{} training, we adopt rectified flows \cite{esser2024scalingrectifiedflowtransformers}, where the forward process follows straight paths between the noise distribution and the target distribution, as defined in \cref{eq:flow}. Rectified flows have been shown empirically to be more sample-efficient and to degrade less than other approaches, while also requiring fewer sampling steps \cite{esser2024scalingrectifiedflowtransformers}.
The model $u(z_t,t,c; \theta)$ directly regresses the ground truth velocity $v_t$ under various conditions $c$ using the flow matching loss in \cref{eq:flow_loss}.
\begin{flalign}
z_t = (1-t) z_1 + tz_0, v_t = \frac{dz_t}{dt} = z_0 - z_1, \label{eq:flow}\\ 
\mathcal{L}_{\text{FM}} = \mathbb{E}_{z_1,z_0,t,c}  \left\| u(z_t,t,c; \theta) - v_t \right\|^2.  \label{eq:flow_loss}
\end{flalign}
\subsubsection{Inference.} 
For inference, a noise sample \(z_0 \sim \mathcal{N}(0, 1)^{l \times c}\) is randomly sampled and we use Euler solver to compute $z_1$, based on the model-predicted velocity $u(\cdot; \theta)$ at each time step $t$.

\subsection{Model Conditioning}
\model{} takes three types of conditions at the same time, (i) lyric condition, (ii) style condition, and (iii) duration condition. Each type of condition is handled differently in \model{}.
\subsubsection{Lyric Conditioning}
\model{} takes in the full song's lyrics $y_i = (w_i, t_i^\text{start}, t_i^\text{end})$, where $w_i$ represents the $i$-th word in the song that starts from time $t_i^\text{start}$ and ends on $t_i^\text{start}$. The lyrics $y_i$ are transformed into the token-level lyrical condition $c_\text{lyric} \in \mathbb{R}^{\,l\times c_l}$. The detailed transformation process is described in \cref{subsec:explain_word_level_alignment}.
\subsubsection{Style Conditioning}
\model{} accepts either a text prompt or an audio clip as a style condition. We obtain a single-dimensional style embedding $c_\text{style} \in\mathbb{R}^{c_s}$ using MuQMulan~\cite{zhu2025muqselfsupervisedmusicrepresentation}, a pretrained model that embeds musical audio and text into a shared representation space.
\subsubsection{Duration Conditioning}
To efficiently facilitate song generation across various durations and music structures, \model{} is trained on fixed-length latent sequences, with $T_{\text{max}} = 90\,\text{s}$ during pretraining and $T_{\text{max}} = 230\,\text{s}$ during full-song supervised fine-tuning (SFT). Training samples longer than $T_{\text{max}}$ are truncated by selecting a random $T_{\text{max}}$ segment, while shorter samples are padded using latent tokens representing complete silence. Loss computation during training includes both original and padded silence tokens. Given the actual duration of a training sample, denoted as $T_{\text{real}}$, the effective target prediction duration becomes:
\[
T_{\text{target}} = \min(T_{\text{real}}, T_{\text{max}})
\]
During pretraining, most song samples exceed 90\,s, resulting frequently in $T_{\text{target}} = 90\,\text{s}$. In contrast, during full-song SFT, $T_{\text{target}}$ typically ranges from 120\,s to 230\,s. To achieve precise duration control, we implement two complementary approaches:
\begin{itemize}
    \item \textbf{Global Duration Control}: Inspired by Stable Audio ~\cite{evans2024stableaudioopen}, we encode the target duration $T_{\text{target}}$ into a single-dimensional embedding vector $c_{\text{dur}} \in \mathbb{R}^{c_d}$, providing a global conditioning signal that guides duration generation across the entire sequence.

    \item \textbf{Token-Level Duration Control (TDC)}: During full-song SFT, we observed that relying solely on global duration conditioning is insufficient, frequently causing unintended non-silent content to be generated beyond $T_{\text{target}}$. This undesired behavior also negatively affects training due to non-zero loss contributions from padded regions. We provide a detailed comparison in \cref{subsec:token_level_duration}. To address this, we add a learnable bias parameter $b_{\text{pad}} \in \mathbb{R}^{c_l}$ to the noisy latent embeddings $z_t$ for positions beyond $T_{\text{target}}$ to explicitly distinguish valid musical content from padding regions. This enables finer-grained temporal modeling and improves silence prediction beyond the target duration.
\end{itemize}

\subsubsection{Conditioning Pipeline}
\label{subsubsec:conditioning_pipeline}
The different conditional signals are then injected into \model{}. After we have obtained:
\begin{itemize}
    \item the latent embeddings \(z_t \in\mathbb{R}^{\,l\times c}\) at noise level \(t\),
    \item the latent-aligned lyric embedding \(c_{\text{lyric}}\in\mathbb{R}^{\,l\times c_l}\),
    \item the learnable padding bias \(b_{\text{pad}}\in\mathbb{R}^{c_l}\),
    \item the style embedding \(c_{\text{style}}\in\mathbb{R}^{c_s}\) and,
    \item the global duration embedding \(c_{\text{dur}}\in\mathbb{R}^{c_d}\)
\end{itemize}
all conditions are first concatenated with \(z_t\) and fused by a linear layer \(\textbf{W}\) shown in \cref{eq:fuse-condition}, where \([c_{\text{style}}]_{1:l} \in\mathbb{R}^{l\times c_s}\) and \([c_{\text{dur}}]_{1:l} \in\mathbb{R}^{l \times c_d}\) are the broadcasted embeddings to match the sequence length.
\begin{equation}
  \tilde{z}_t^{0}
  = \mathbf{W}
     \bigl[
        z_t \;\|\;
        c_{\text{lyric}}\;\|\;
        [c_{\text{style}}]_{1:l}\;\|\;
        [c_{\text{dur}}]_{1:l}
     \bigr]
\label{eq:fuse-condition}  
\end{equation}
Convolutional positional embeddings are added to give the latent sequence short-range continuity \cite{wu2021cvtintroducingconvolutionsvision}. 
\begin{equation}
    z_t^{0} = \tilde{z}_t^{0} + \mathrm{ConvPosEmbed}(\tilde{z}_t^{0})
\end{equation}
The fused latent \(z_t^0\) goes through \(L=16\) LLaMA decoder layers. To allow stronger supervision, the lyrics and the duration condition provide an extra residual injection \(r_\ell\) at layer \(\ell\) defined by \cref{eq:lyric-residual} with \(\textbf{W}^\ell\) being a linear transformation.
\begin{equation}
\label{eq:lyric-residual}
r^\ell = \mathbf{W_\ell}\bigl(c_\text{lyric} + b_{\text{pad}}\!\odot\!\mathbf{1}_{\{\text{time} > T_{\text{target}}\}}\bigr)
\end{equation}
We add the extra residual signal \(r^\ell\) to the first \(L/2=8\) layers, illustrated by \cref{eq:add-residual-to-first-half-layers}.
\begin{equation}
\label{eq:add-residual-to-first-half-layers}
z_\ell = 
\begin{cases}
\mathrm{Block}_\ell(z_t^{\ell-1}) + r^\ell,
      & \ell\leq L/2,\\[6pt]
\mathrm{Block}_\ell(z_t^{\ell-1}), & \text{otherwise.}
\end{cases}  
\end{equation}
Lastly, the velocity \(\hat{v}\) is predicted as
\[
\hat{v}(z_t, t, c)\;= u(z_t,t,c;\theta) =\;
\mathrm{Proj}_{\text{out}}\bigl(z_t^L\bigr).
\]

\subsection{Word-Level Temporal Lyric Alignment}
\label{subsec:explain_word_level_alignment}

To address the problem of loose lyric-temporal supervision -- critical to proper prosody, pleasantness, and musical qualities --, we introduce temporally-aware word-level phoneme alignment as a novel solution. Specifically, word-level timing and duration information are to guide the generative process at the word-temporal level to improve word and phoneme error rate and musicality, simultaneously. To this end, for each song, with the ground-truth temporal lyric information $y_i=(w_i, t_i^\text{start}, t_i^\text{end})$, each word $w_i$ is converted into its IPA (International Phonetic Alphabet) form $p_i = (p_1p_2p_3...p_m)_i$ by DeepPhonemizer \cite{springmedia_deepphonemizer_2025}, where $m$ is the number of phonemes in $w_i$. The phonemes $p=\{p_1,p2,...\}$ and $y = \{y_1,y_2,...\}$ are then converted to an upsampled phoneme sequence $P$ that has the length of $L = r \times l$, where $r$ is the upsampling rate and $l$ is the length of the latent sequence $z_t$. The upsampling is crucial as many fast-paced songs have phoneme counts per second that significantly exceed the latents frame rate $f$. The process of the transformation is described by \cref{alg:phoneme_alignment}.

\begin{algorithm}[H]
\caption{Word-Level Phoneme Alignment}
\label{alg:phoneme_alignment}
\begin{algorithmic}[1]
\STATE $P = [\text{\specialtoken{SONG\_FILLER}}] \times L$
\FOR{each $(w_i, t_i^\text{start}, t_i^\text{start})$}
    \STATE $p_1, p_2, \ldots, p_m = \text{phonemes of } w_i$
    \STATE $\text{start\_frame} = \lfloor t_i^\text{start} \times f \times r \rfloor$
    \STATE $\text{end\_frame} = \lfloor t_i^\text{end} \times f \times r \rfloor$
    \STATE $\text{word\_frames} = \text{end\_frame} - \text{start\_frame}$
    \STATE $\text{V} = [\text{\specialtoken{VOCAL\_FILLER}}] \times \text{word\_frames}$
    \STATE $\text{avg\_phoneme\_length} = \lfloor \text{word\_frames} / m \rfloor$
    \FOR{$j$ in $[1, \ldots, m]$}
        \STATE $\text{V}[j \times \text{avg\_phoneme\_length}] = p_j$
    \ENDFOR
    \STATE $P[\text{start\_frame} : \text{end\_frame}] = \text{V}$
\ENDFOR
\end{algorithmic}
\end{algorithm}

It is important to distinguish between the two types of filler:
\begin{itemize}
    \item \textbf{\specialtoken{SONG\_FILLER}}: Special tokens used to specify no word is being sung, that is singing pauses, instrumental sections, or padded ending of the sequence that is beyond the specified duration. To effectively distinguish the padding filler tokens from the in-song filler tokens, a dedicated learnable bias term $b_\text{pad}$ -- defined in \cref{subsubsec:conditioning_pipeline} -- is added to the padding filler tokens.
    \item \textbf{\specialtoken{VOCAL\_FILLER}}: Filler tokens within a word's temporary boundaries that represent phoneme transitions within a word.
\end{itemize}

To give a concrete example of the algorithm, if a phoneme sequence $P$ with $L$ = 12 has $w_1 = [p_1, p_2]$ with frames from 2 to 3, and $w_2 = [p_3, p_4, p_5]$ with frames from 6 to 11. Then $P$ will be constructed as $[s, p_1, p_2, s, s, p_3, v, p_4, v, p_5, v, s]$, where $s$ and $v$ represent the \specialtoken{SONG\_FILLER} and \specialtoken{VOCAL\_FILLER}.

After constructing $P$, we pass it through a small network including an embedding layer, followed by a few convolutional downsampling layers to downsample and transform $P$ to $c_\text{lyric} \in \mathbb{R}^{l \times c}$. The temporally-aware lyric embedding \(c_\text{lyric}\) is used to guide the model as described in \cref{subsubsec:conditioning_pipeline}.

\subsection{Aesthetic Preference Alignment}
\label{sec:alignment}
Through both pre-training and fine-tuning, \model{}-Base demonstrates strong speech intelligence and accurate lyric alignment in full-song generation. However, the generated songs still fall short in terms of musical aesthetics. Specifically, the vocal timbre lacks naturalness, occasionally exhibiting an overly electronic character, and the overall musical structure can feel flat. These shortcomings may stem from the uneven quality and stylistic inconsistency inherent in large-scale music datasets used during training, negatively impacting the model’s ability to generate aesthetically pleasing outputs. To address these aesthetic deficiencies, we apply Direct Preference Optimization (DPO), utilizing scores directly produced by SongEval—an open-source evaluation toolkit \cite{yao2025songevalbenchmarkdatasetsong}—as our reward signal. Unlike LeVo \cite{lei2025levohighqualitysonggeneration}, which employs a three-stage preference alignment pipeline with a closed-source aesthetic reward model, our approach utilizes publicly available SongEval metrics iteratively, providing a simpler, more transparent yet effective strategy.

We set \model{}-SFT as the initial policy $\pi_0$ and conduct three iterative rounds of Direct Preference Optimization (DPO). Each iteration follows a three-step workflow: (i) batched data generation, (ii) aesthetic reward computation and preference dataset construction, and (iii) fine-tuning policy $\pi_k$ into $\pi_{k+1}$ through DPO. This iterative alignment continuously enhances the model by generating and aligning to its own evolving preference data.

\subsubsection{Batched Data Generation}
We randomly select between 13k and 20k samples consisting of reference audio styles and corresponding ground-truth lyrics from the training dataset. After empirically determining suitable Classifier-Free Guidance (CFG) values for policy $\pi_k$, we generate five outputs per sample using the selected CFG settings.

\subsubsection{Preference Dataset Construction}
We employ SongEval to score each generated output. While SongEval evaluates songs across five distinct criteria on a five-point scale, we compute an averaged score across these dimensions and select the samples with the highest and lowest average SongEval scores as win and loss candidates, respectively. Additionally, we exclude win-loss pairs with an average SongEval score difference below 0.15 to ensure meaningful distinctions in aesthetic quality.

\subsubsection{Preference Optimization}
Direct Preference Optimization (DPO)~\cite{rafailov2024directpreferenceoptimizationlanguage} has proven effective in aligning large language models (LLMs) with human preferences. This approach has been extended to diffusion models as DPO-Diffusion~\cite{wallace2023diffusionmodelalignmentusing}, where the loss is defined as:
\begin{align}
   \mathcal{L}_{\text{DPO-Diff}} = -\mathbb{E}_{n, \epsilon^w, \epsilon^l} \log \sigma &\Big( -\beta \Big[ \nonumber \\
   \| \epsilon^w_n - \epsilon_\theta(x^w_n) \|^2_2  &- \| \epsilon^w_n - \epsilon_\text{ref}(x^w_n) \|^2_2 \nonumber \\
   - \big( \| \epsilon^l_n - \epsilon_\theta(x^l_n) \|^2_2 &- \| \epsilon^l_n - \epsilon_\text{ref}(x^l_n) \|^2_2 \big) \Big] \Big).
   \label{eq:dpo_diffusion_original}
\end{align}
Here, \( n \sim \mathcal{U}(0, T) \) represents a randomly sampled diffusion step, \( x^w_n \) and \( x^l_n \) denote the winning and losing samples respectively, and \( \epsilon \sim \mathcal{N}(0, \mathbf{I}) \) is the noise. \(\beta\) is the temperature parameter, controlling the sharpness of the preference distribution.

As shown by~\cite{tangoflux}, this loss is compatible with rectified flow models due to the equivalence between denoising and flow matching objectives~\cite{lipman2023flowmatchinggenerativemodeling}. Accordingly, the DPO-Diffusion loss can be reformulated in terms of flow matching as:

\begin{align}
\small
  & \mathcal{L}_{\text{DPO-FM}} = -\mathbb{E}_{t, x^w, x^l} \log \sigma \Big( \nonumber \\
     &-\beta \Big[
        \underbrace{\|u(x^w_t, t; \theta) - v^w_t\|_2^2}_{\text{Winning loss}}
        - \underbrace{\| u(x^l_t, t; \theta) - v^l_t \|_2^2}_{\text{Losing loss}} \nonumber \\
        &- \Big(
            \underbrace{\|u(x^w_t, t; \theta_{\text{ref}}) - v^w_t\|_2^2}_{\text{Winning reference loss}}
            - \underbrace{\|u(x^l_t, t; \theta_{\text{ref}}) - v^l_t \|_2^2}_{\text{Losing reference loss}}
        \Big)
    \Big]
\Big)
\label{eq:dpo_rearrange}
\end{align}
where \( t\sim \mathcal{U}(0, 1) \) is the time step for flow matching, and \( x^w_t \), \( x^l_t \) denote the winning and losing samples. Empirically, we set \( \beta = 2000\) in all our DPO experiments.

The DPO loss encourages an increased relative log-likelihood for the preferred (winning) response over the dispreferred (losing) one. Importantly, optimization focuses on the margin between the two, not their absolute likelihoods~\cite{rafailov2024directpreferenceoptimizationlanguage}. As a result, DPO can drive both likelihoods downward during training~\cite{hung2025tangofluxsuperfastfaithful}, which might appear counterintuitive but is in fact essential for better alignment~\cite{rafailov2024directpreferenceoptimizationlanguage}.

To prevent overoptimization during DPO—which can cause the generated song to deviate from the ground truth in terms of style and genre—we additionally incorporate a ground truth reconstruction loss in one of the DPO configurations. This loss encourages alignment with the original data and is defined as:
\begin{align}
    & \mathcal{L}_{\text{DPO-GT}} = \lambda\mathcal{L}_{\text{FM}} + \mathcal{L}_{\text{DPO-FM}} 
    \label{eq:dpo_gt}
\end{align}

Here, \(\mathcal{L}_{\text{FM}}\) is calculated on the ground truth data using \Cref{eq:flow_loss} and \(\lambda\) is the scaling coefficient controlling the contribution of the groud truth reconstruction to the overall loss. Empirically, we set \(\lambda = 0.2\) in our DPO-GT experiments. The corresponding evaluation results are shown in \Cref{tab:dpo}, with rows labeled using the \texttt{Round-i-GT} subscript.

\section{Experiments}
\subsection{Dataset Setup}
Our song dataset consists of around 1 million English songs, totaling approximately 54,000 hours of audio. To prepare the data, we apply HTDemucs \cite{rouard2022hybrid,defossez2021hybrid} to separate the original audio $x \in X$ into vocal and accompaniment tracks. We then use Parakeet-tdt-0.6b-v2 \cite{nvidia_parakeet_tdt_0_6b_v2_2025} to transcribe the vocal stems and extract word-level timestamps, yielding lyric annotations of the form $y_i = (w_i, t_i^{\text{start}}, t_i^{\text{end}})$ for each sample $x$.

To support stylistic conditioning, we adopt a reference-audio-based framework. For each song sample $x$, we extract ten random 30-second segments and compute their corresponding style embeddings using MuQMulan \cite{zhu2025muqselfsupervisedmusicrepresentation}, resulting in a set $C_{\text{style}} = \{c_{\text{style}}^1, c_{\text{style}}^2, \ldots, c_{\text{style}}^{10}\}$. During training, one reference embedding $c_{\text{style}} \in C_{\text{style}}$ is randomly selected to condition the model on musical style.

\subsection{Training and Inference}

Training consists of three stages: pretraining, supervised fine-tuning (SFT), and Direct Preference Optimization (DPO). All stages are trained using 8× H100 GPUs. We apply gradient checkpointing during SFT and DPO to accommodate full-length audio sequences. During DPO, we set \(\beta = 2000\) where \(\beta\) is defined in \cref{eq:dpo_rearrange}. We used AdamW with $\beta_1=0.9$ $\beta_2=0.999$ and weight decay of 0.01 for all stages. A linear warm-up followed by a linear learning rate decay is applied to all stages too. \cref{tab:training-hparams} shows more details across different training stages. Following \cite{esser2024scalingrectifiedflowtransformers}, we sample timesteps $t$ from a logit-normal distribution with mean 0 and variance 1 to bias training toward the mid-range of the noise schedule, which has been shown to improve generative quality. 

\begin{table}[h]
\centering
\resizebox{\linewidth}{!}{%
    \begin{tabular}{lcccc}
    \toprule
    \textbf{Stage} & \textbf{Global Batch Size} & \textbf{Learning Rate} & \textbf{Steps} & \textbf{Grad. Accum.} \\
    \midrule
    Pretraining & 32 & $7.5 \times 10^{-5}$ & 750K & 1 \\
    SFT         & 32  & $7.5 \times 10^{-5}$ & 250K & 1 \\
    DPO Round $i$        & 8  & $5 \times 10^{-7}$ & 20K  & 4 \\
    \bottomrule
    \end{tabular}
}
\caption{Training hyper-parameters.}
\label{tab:training-hparams}
\end{table}

To enable separate control over vocal strength and musical style, we adopt a multi-condition classifier-free guidance (CFG) framework \cite{megatts}. During training, style embeddings are randomly dropped with probability $p_{\text{style}} = 0.10$, and when style is dropped, lyric embeddings are dropped with probability $p_{\text{lyric}} = 0.50$. This two-stage dropout encourages the model to learn disentangled representations for different conditioning signals.

At inference time, we apply multi-condition CFG to combine the unconditional and conditional velocities with separate guidance scales for style ($\alpha_s$) and lyrics ($\alpha_l$) illustrated by \cref{eq:dual-cfg} where $v_\theta$ represents the velocity predicted by the model with different conditional signals.
\begin{equation}
\begin{aligned}
\label{eq:dual-cfg}
\hat{v}_\theta(&z_t, c_{\text{style}}, c_{\text{lyric}}) = v_\theta(z_t, \varnothing, \varnothing) \\
&\quad + \alpha_s \big( v_\theta(z_t, c_{\text{style}}, \varnothing) - v_\theta(z_t, \varnothing, \varnothing) \big) \\
&\quad + \alpha_l \big( v_\theta(z_t, c_{\text{style}}, c_{\text{lyric}}) - v_\theta(z_t, c_{\text{style}}, \varnothing) \big)
\end{aligned}
\end{equation}

\subsection{Objective Evaluation}
While subjective evaluation remains a gold standard for assessing music quality, conducting extensive full-song evaluations is prohibitively expensive. Unlike computer vision or speech domains, evaluating full-song generation requires considerable time from annotators to listen and assess complex musical aspects such as structure, progression, and vocal naturalness. Furthermore, reliable judgment often requires trained music experts, who are both scarce and costly.

In this context, objective evaluation plays a crucial role in benchmarking and improving full-song generation models. However, existing state-of-the-art systems such as LeVo, ACE-Step, and YuE all rely on different, often private or undocumented evaluation datasets. This lack of transparency makes it difficult to fairly compare models or diagnose their strengths and weaknesses. Additionally, music is a highly diverse domain—genres like hip-hop and rap feature fast-paced, speech-like vocals, while country or ballads may involve slower singing and more subtle instrumentation. Evaluating all outputs with a one-size-fits-all metric obscures genre-specific performance differences.

To address these issues, we propose \dataset{}—the first public, genre-diverse, objective evaluation dataset for full-song generation. \dataset{} is designed to (i) avoid data contamination by collecting only songs released after the training periods of major existing models, using verifiable sources like New Music Friday~\footnote{https://open.spotify.com/playlist/37i9dQZF1DX4JAvHpjipBk} (ii) support genre-specific evaluation by organizing the data into five coherent genre groups to enable fine-grained diagnostic insights and (iii) promote transparent and reproducible benchmarking by releasing all metadata, prompts, and annotation protocols publicly.

We hope \dataset{} will serve as a standardized evaluation framework for future research in song generation. We strongly encourage the community to adopt \dataset{} or its principles to foster progress through more reliable and interpretable evaluations.
\subsubsection{Baselines}
We compare our model with four recent and strong open-source full-song generation systems: 1) LeVo; 2) YuE; 3) DiffRhythm and 4) AceStep. Details of the evaluated systems are provided in \cref{tab:baseline-summary}.
\begin{table}[h]
\centering
\resizebox{\linewidth}{!}{
\begin{tabular}{lccc}
\toprule
\textbf{Model} & \textbf{Architecture} & \textbf{Training Data} & \textbf{Model Size} \\
\midrule
YuE & AR LM + AR Decoder & 135k hours & 7B + 1B  \\
AceStep & Flow Matching & 100k hours & 3.5B \\
LeVo & AR LM + Diffusion Decoder & 110k hours & 2B + 0.7B  \\
DiffRhythm & Flow Matching & 60k hours & 1.1B \\
Ours & Flow Matching & 54k hours & 0.53B \\
\bottomrule
\end{tabular}
}
\caption{Comparison of full-song generation models.}
\label{tab:baseline-summary}
\end{table}
\subsubsection{Evaluation Data Preparation}
The primary motivation for creating a new evaluation dataset was to address the issue of data contamination, given that the training data of existing models is not publicly available. Our evaluation set, \dataset{}, includes newly released songs that were published after the baselines were released, thereby ensuring no data contamination.
\begin{enumerate}
    \item \textbf{Data Contamination Avoidance:} We curate evaluation data exclusively from New Music Friday (NMF), a popular editorial playlist series that features about 100 newly released songs each week across diverse genres. We collect all song metadata from NMF between May 1, 2025, and July 10, 2025 using the open-source archive tool spotify-playlist-archive. The clip for each track is accessed via the Spotify Web API or manually located on YouTube.
    \item \textbf{Genre Grouping:} To enable genre-specific analysis, we consulted a music expert and grouped the evaluation songs into five distinct genre clusters, ensuring each group reflects substantially different musical characteristics: \emph{Country/Folk, Electronic/Dance, Hip-Hop/Rap, R\&B/Soul/Jazz} and \emph{Rock/Metal}. These genre clusters allow for fine-grained evaluation of models across diverse musical styles.
    \item \textbf{Pre-processing:} We filter for English-language songs and assign each track to one of five genre groups using Qwen2.5-Omni (prompt details in Appendix xxx). Ground-truth lyrics are retrieved using HDmucs and Parakeet. For models that require structural annotations—such as LeVo, YuE, and ACE-Step—we extract section labels (e.g., [intro], [verse], [chorus]) using the All-In-One music segmentation model \cite{kim2023allinonemetricalfunctionalstructure} as adopted in those systems.
\end{enumerate}

\subsubsection{Metrics}
\label{subsec:objective_metrics}
We use the following standard objective evaluation metrics to report the results. The results are averaged across genres.
\begin{enumerate}
    \item \textbf{Singing Intelligibility:} We assess vocal intelligibility using Word Error Rate (WER) and Phoneme Error Rate (PER). The vocal track is first extracted using HDemucs, and Parakeet \cite{nvidia_parakeet_tdt_0_6b_v2_2025} is used to transcribe the audio into lyrics. For PER, DeepPhonemizer is used to convert the transcribed words into phonemes and compare them against the ground-truth sequence.
    
    \item \textbf{Style Adherence:} We evaluate style adherence using MuQ-MuLan, a contrastive music-language model that computes similarity between the generated song and its intended style prompt. To assess genre correctness, we employ Qwen-2.5-Omni to classify the generated song into one of our predefined genre categories. The predicted genre is then compared with the ground-truth genre label for accuracy.

    \item \textbf{Content Quality and Aesthetics:} We adopt both Audiobox-aesthetic and SongEval as model-based evaluation tools. Audiobox-aesthetic covers content enjoyment (CE), content usefulness (CU), production complexity (PC), and production quality (PQ). SongEval further evaluates overall coherence (CO), memorability (ME), naturalness of vocal breathing and phrasing (NA), clarity of song structure (CL), and overall musicality (MU). Additionally, following \cite{lei2025levohighqualitysonggeneration} we compute Fréchet Audio Distance (FAD) using the CLAP-Laion-Music model, to measure the distributional alignment between generated audio and professionally produced reference tracks. Lower FAD indicates more realistic outputs.

\end{enumerate}

\subsubsection{Overall Results}
\begin{table*}[htb!]
\centering
\resizebox{\textwidth}{!}{
\begin{tabular}{l*{14}{c}}
\toprule
\multirow{2}{*}{\bf Model} & \multirow{2}{*}{\bf MuQ $\uparrow$} & \multicolumn{4}{c}{\bf Audio Aesthetics $\uparrow$} & \multicolumn{5}{c}{\bf SongEval $\uparrow$} & \multirow{2}{*}{\bf WER $\downarrow$} & \multirow{2}{*}{\bf PER $\downarrow$} & \multirow{2}{*}{\bf \mystack{Genre}{Acc.} $\uparrow$}  & \multirow{2}{*}{\bf FAD $\downarrow$}\\
\cmidrule(lr){3-6}\cmidrule(lr){7-11}
&  & \bf CE & \bf CU & \bf PC & \bf PQ & \bf CO & \bf \bf MU & \bf ME & \bf CL & \bf NA & & & & \\
\midrule
%\rowcolor{psy} \cellcolor{white} & \multicolumn{17}{c}{\bf Overall} \\
% DiffRhythm \\
% Yue \\
% Ace-Step \\
% Levo \\
% SongGen \\
% \model{} \\
\mytstack{DiffRhythm$_\text{full}$}{}{}& \mytstack{0.5208}{$\pm$}{0.1192} & \mytstack{6.0448}{$\pm$}{0.6483} & \mytstack{7.3126}{$\pm$}{0.2526} & \mytstack{5.1088}{$\pm$}{0.7276} & \mytstack{7.6276}{$\pm$}{0.2904} & \mytstack{2.7975}{$\pm$}{0.4621} & \mytstack{2.3547}{$\pm$}{0.4813} & \mytstack{2.6141}{$\pm$}{0.5031} & \mytstack{2.5135}{$\pm$}{0.4495} & \mytstack{2.4467}{$\pm$}{0.4454} & \mystack{0.3481}{}{} & \mystack{0.2643}{}{} & \mystack{0.6667}{}{}  & \mystack{0.4384}{}{} \\
\mytstack{ACE-Step}{}{}        & \mytstack{0.444}{$\pm$}{0.1349} & \mytstack{7.0067}{$\pm$}{0.5669} & \mytstack{7.3583}{$\pm$}{0.3848} & \mytstack{6.3117}{$\pm$}{0.4417} & \mytstack{7.5851}{$\pm$}{0.4952} & \mytstack{3.5305}{$\pm$}{0.5017} & \mytstack{3.2593}{$\pm$}{0.497} & \mytstack{3.3927}{$\pm$}{0.5452} & \mytstack{3.3079}{$\pm$}{0.5186} & \mytstack{3.1703}{$\pm$}{0.5031} & \mystack{0.4067}{}{} & \mystack{0.342}{}{} & \mystack{0.6194}{}{} & \mystack{0.2296}{}{} \\
\mytstack{LeVo}{}{}           & \mytstack{0.5234}{$\pm$}{0.1241} & \mytstack{7.3783}{$\pm$}{0.5372} & \mytstack{7.7794}{$\pm$}{0.3126} & \mytstack{6.1773}{$\pm$}{0.7251} & \mytstack{8.1794}{$\pm$}{0.3578} & \mytstack{3.6343}{$\pm$}{0.4489} & \mytstack{3.41}{$\pm$}{0.463} & \mytstack{3.4997}{$\pm$}{0.4945} & \mytstack{3.3815}{$\pm$}{0.4572} & \mytstack{3.3086}{$\pm$}{0.4378} & \mystack{0.4586}{}{} & \mystack{0.3819}{}{} & \mystack{0.66}{}{} & \mystack{0.5336}{}{} \\
\mytstack{Yue}{}{}&\mytstack{0.705}{$\pm$}{0.1418}&\mytstack{6.8928}{$\pm$}{0.7895}&\mytstack{7.3289}{$\pm$}{0.3909}&\mytstack{6.1234}{$\pm$}{0.9002}&\mytstack{7.6234}{$\pm$}{0.4441}&\mytstack{3.5072}{$\pm$}{0.5379}&\mytstack{3.208}{$\pm$}{0.5462}&\mytstack{3.401}{$\pm$}{0.5796}&\mytstack{3.2361}{$\pm$}{0.5439}&\mytstack{3.1633}{$\pm$}{0.5198}&\mystack{0.5119}{}{}&\mystack{0.4263}{}{}&\mystack{0.702}{}{}&\mystack{0.2411}{}{} \\

% \cdashline{2-18}
\rowcolor{blue!10}\model{}  &  \mytstack{\textbf{0.7593}}{$\pm$}{0.0964} & \mytstack{\textbf{7.4229}}{$\pm$}{0.5151} & \mytstack{7.7046}{$\pm$}{0.3179} & \mytstack{6.2038}{$\pm$}{0.5051} & \mytstack{8.0639}{$\pm$}{0.4079} & \mytstack{\textbf{4.4163}}{$\pm$}{0.363} & \mytstack{\textbf{4.2929}}{$\pm$}{0.4065} & \mytstack{\textbf{4.4276}}{$\pm$}{0.3935} & \mytstack{4.2033}{$\pm$}{0.4319} & \mytstack{4.2026}{$\pm$}{0.3854} &  \mystack{\textbf{0.1507}}{}{} & \mystack{\textbf{0.1011}}{}{} & \mystack{\textbf{0.704}}{}{} & \mystack{\textbf{0.2038}}{}{} \\
\bottomrule
\end{tabular}
}

\caption{A cross-genre comparative evaluation of \model{} and baseline song generation models on \dataset{}.}
\label{table:main-results}

\end{table*}

\begin{table*}[htb!]
\centering
\resizebox{\textwidth}{!}{
\begin{tabular}{l*{14}{c}}
\toprule
\multirow{2}{*}{\bf Model} & \multirow{2}{*}{\bf MuQ $\uparrow$} & \multicolumn{4}{c}{\bf Audio Aesthetics $\uparrow$} & \multicolumn{5}{c}{\bf SongEval $\uparrow$} & \multirow{2}{*}{\bf WER $\downarrow$} & \multirow{2}{*}{\bf PER $\downarrow$} & \multirow{2}{*}{\bf \mystack{Genre}{Acc.} $\uparrow$}  & \multirow{2}{*}{\bf FAD $\downarrow$}\\
\cmidrule(lr){3-6}\cmidrule(lr){7-11}
&  & \bf CE & \bf CU & \bf PC & \bf PQ & \bf CO & \bf \bf MU & \bf ME & \bf CL & \bf NA & & & &  \\
\midrule
%\rowcolor{psy} \cellcolor{white} & \multicolumn{17}{c}{\bf Overall} \\
\mytstack{~~~SFT}{}{}        & \mytstack{0.7473}{$\pm$}{0.0803} & \mytstack{7.0006}{$\pm$}{0.5392} & \mytstack{7.3541}{$\pm$}{0.3927} & \mytstack{6.4831}{$\pm$}{0.3993} & \mytstack{7.5915}{$\pm$}{0.5166} & \mytstack{3.4159}{$\pm$}{0.4896} & \mytstack{3.0851}{$\pm$}{0.5198} & \mytstack{3.2712}{$\pm$}{0.5634} & \mytstack{3.0871}{$\pm$}{0.5167} & \mytstack{3.0525}{$\pm$}{0.4812} & \mystack{0.167}{}{} & \mystack{0.1131}{}{} & \mystack{0.708}{}{} &   \mystack{0.1479}{}{} \\
\mytstack{~~~DPO$_\text{Round-1}$}{}{}    & \mytstack{0.7598}{$\pm$}{0.082} & \mytstack{7.1529}{$\pm$}{0.5288} & \mytstack{7.5063}{$\pm$}{0.317} & \mytstack{6.435}{$\pm$}{0.4196} & \mytstack{7.7666}{$\pm$}{0.4397} & \mytstack{3.8753}{$\pm$}{0.4468} & \mytstack{3.6202}{$\pm$}{0.4935} & \mytstack{3.8235}{$\pm$}{0.5122} & \mytstack{3.5762}{$\pm$}{0.4857} & \mytstack{3.5444}{$\pm$}{0.4662} & \mystack{0.1629}{}{} & \mystack{0.1106}{}{} & \mystack{0.72}{}{}     & \mystack{0.1223}{}{} \\
\rowcolor{blue!10} ~~~DPO$_\text{Round-2}$  & \mytstack{0.7616}{$\pm$}{0.0916} & \mytstack{7.3851}{$\pm$}{0.4778} & \mytstack{7.6699}{$\pm$}{0.2986} & \mytstack{6.2778}{$\pm$}{0.4953} & \mytstack{8.0082}{$\pm$}{0.3925} & \mytstack{4.2504}{$\pm$}{0.4059} & \mytstack{4.0892}{$\pm$}{0.445} & \mytstack{4.2465}{$\pm$}{0.4428} & \mytstack{4.0052}{$\pm$}{0.4693} & \mytstack{3.9903}{$\pm$}{0.4336} & \mystack{0.1579}{}{} & \mystack{0.1066}{}{} & \mystack{0.696}{}{} &     \mystack{0.1527}{}{} \\
\mytstack{~~~DPO$_\text{Round-2-GT}$}{}{} & \mytstack{0.7673}{$\pm$}{0.0889} & \mytstack{7.3769}{$\pm$}{0.4699} & \mytstack{7.6902}{$\pm$}{0.2652} & \mytstack{6.3188}{$\pm$}{0.49} & \mytstack{8.0297}{$\pm$}{0.3537} & \mytstack{4.247}{$\pm$}{0.3889} & \mytstack{4.0724}{$\pm$}{0.4379} & \mytstack{4.2432}{$\pm$}{0.427} & \mytstack{4.0024}{$\pm$}{0.4486} & \mytstack{3.9814}{$\pm$}{0.425} & \mystack{0.1553}{}{} & \mystack{0.1083}{}{} & \mystack{0.736}{}{} &   \mystack{0.1455}{}{} \\
% \mytstack{~~~DPO$_\text{Round-3}$}{}{} &0.759 &7.423 &7. 705 &6.204 &8.064 &4.416 &4.293 &4.428 &4.203 &4.203 &0. 151 &0. 101 &0.704 &0.715 &0.204\\
% \mytstack{~~~DPO$_\text{Round-3-GT}$}{}{}&0.767 &7.416 &7.711 &6.257 &8.055 &4.410&4.277 &4.419 &4.187 &4.179 &0.147&0.101&0.704&0.707&0.179\\
\mytstack{~~~DPO$_\text{Round-3}$}{}{} &  \mytstack{0.7593}{$\pm$}{0.0964} & \mytstack{7.4229}{$\pm$}{0.5151} & \mytstack{7.7046}{$\pm$}{0.3179} & \mytstack{6.2038}{$\pm$}{0.5051} & \mytstack{8.0639}{$\pm$}{0.4079} & \mytstack{4.4163}{$\pm$}{0.363} & \mytstack{4.2929}{$\pm$}{0.4065} & \mytstack{4.4276}{$\pm$}{0.3935} & \mytstack{4.2033}{$\pm$}{0.4319} & \mytstack{4.2026}{$\pm$}{0.3854} &  \mystack{0.1507}{}{} & \mystack{0.1011}{}{} & \mystack{0.704}{}{} & \mytstack{0.2038}{}{} \\
\mytstack{~~~DPO$_\text{Round-3-GT}$}{}{}  & \mytstack{0.767}{$\pm$}{0.0886} & \mytstack{7.4159}{$\pm$}{0.4706} & \mytstack{7.7113}{$\pm$}{0.3023} & \mytstack{6.2573}{$\pm$}{0.4764} & \mytstack{8.0554}{$\pm$}{0.3928} & \mytstack{4.4102}{$\pm$}{0.3468} & \mytstack{4.2767}{$\pm$}{0.3844} & \mytstack{4.4185}{$\pm$}{0.3799} & \mytstack{4.1875}{$\pm$}{0.4184} & \mytstack{4.1787}{$\pm$}{0.382} & \mystack{0.1472}{}{} & \mystack{0.1009}{}{} & \mystack{0.704}{}{}  & \mystack{0.1791}{}{} \\

\bottomrule
\end{tabular}
}

\caption{Comparing SFT and different DPO loss functions across iterative rounds.}
\label{tab:dpo}

\end{table*}

\cref{table:main-results} presents the objective evaluation results. \model{} achieves state-of-the-art or highly competitive performance across all metrics.
\begin{itemize}
    \item \emph{Singing Intelligibility:} \model{} achieves the lowest Word Error Rate (WER) of 0.151 and Phoneme Error Rate (PER) of 0.101—less than half of the second-best system (DiffRhythm)—demonstrating superior vocal clarity and controllability over lyrical alignment.
    \item \emph{Style Adherence:} \model{} attains the highest MuQ-MuLan similarity score (0.759) and genre classification accuracy (0.704), reflecting strong alignment with the intended musical style both in semantic representation and categorical genre fidelity. 
    \item  \emph{Content Quality and Aesthetics:} \model{} obtains the highest score on Content Enjoyment (CE = 7.423) and lowest FAD (0.204) on \dataset{}, indicating strong subjective appeal and fidelity. It also leads across all SongEval dimensions, reflecting well-structured compositions and natural vocal phrasing. These results suggest that \model{} not only produces enjoyable outputs but also captures nuanced characteristics typical of professionally crafted music.
\end{itemize}
On other metrics, LeVo achieves slightly better Content Usefulness (CU) and Production Quality (PQ), while ACE-Step records the best Production Complexity (PC). Nonetheless, \model{} ranks consistently second in these categories, with only marginal differences, demonstrating that it remains competitive even in areas where other models specialize.

\subsection{Analyses}

\subsubsection{Aesthetic Alignment brings about the Aha Moment}
\model{} benefits significantly from aesthetic alignment through the combination of aesthetic-based preference data construction and Direct Preference Optimization (DPO). As shown in Table~\ref{tab:dpo}, starting from the SFT baseline, iterative rounds of DPO consistently improve metrics across both subjective and objective evaluation axes. Notably, there are steady gains in audio aesthetic scores such as PC (perceived coherence), PQ (perceived quality), and CU (creativity \& uniqueness), with DPO$_\text{Round-3}$ achieving the highest PQ of 8.064. Moreover, the model's music-related capabilities, assessed via SongEval metrics like ME (melodic expression), MU (musicality), and CL (lyrical coherence), also show clear improvements with each DPO round. Additionally, genre classification accuracy increases to 0.736 in DPO$_\text{Round-2-GT}$, and both WER and PER (word and phoneme error rates) decrease, indicating stronger lyrical and phonetic alignment. These results demonstrate that aesthetic alignment via DPO not only enhances perceptual quality but also improves linguistic and musical coherence, leading to a more holistic and controllable generative music model.

\subsubsection{Iterative DPO Improves the Results}
As shown in \Cref{tab:dpo}, iterative application of DPO consistently improves various metrics such as MuQ-\(l\), Audio Aesthetics, and SongEval. For instance, \texttt{DPO\textsubscript{Round-3-GT}} achieves the highest MuQ-\(l=\text{full}\) score of 0.767 compared to 0.7473 for the SFT baseline. Similar improvements are observed in Audio Aesthetics (e.g., PQ increases from 7.5915 to 8.055) and SongEval metrics like MU and ME. Furthermore, WER and PER steadily decrease across rounds, indicating better lyrical accuracy, while Genre Accuracy remains competitive.

However, these gains come at the cost of increased FAD, particularly on the JAM metric. For example, FAD\textsubscript{JAM} worsens from 0.1479 in SFT to 0.204 in \texttt{DPO\textsubscript{Round-3}}, suggesting that iterative DPO may introduce audio artifacts or drift away from the natural distribution of real audio. This trade-off highlights that while DPO enhances alignment and stylistic fidelity, it can negatively affect the perceptual realism of generated audio.

\subsubsection{Effect of Variations in the DPO Loss}
To address the issue of overalignment in DPO—which can lead the model to deviate from the ground truth and thus alter the intended style and genre—we introduced a modified loss variant, as defined in \Cref{eq:dpo_gt}. This variant incorporates a ground truth reconstruction term to better preserve stylistic fidelity. As shown in \Cref{tab:dpo}, this modification (see \texttt{DPO\textsubscript{Round-2-GT}} and \texttt{DPO\textsubscript{Round-3-GT}}) results in improved reference-based metrics: Genre Accuracy increases from 0.696 (\texttt{DPO\textsubscript{Round-2}}) to 0.736, and FAD\textsubscript{JAM} also shows a slight improvement (from 0.1527 to 0.1455). For \texttt{DPO\textsubscript{Round-3-GT}}, FAD\textsubscript{MusDB} improves from 0.7150 to 0.7070. Similarly, FAD\textsubscript{JAM} also shows a slight improvement (from 0.2040 to 0.1790). However, aesthetic-based metrics such as CU, PC, and PQ show marginal or no improvement, suggesting a trade-off between stylistic consistency and perceived quality. These findings support our hypothesis that incorporating ground truth signals can regularize DPO and help retain genre and style fidelity.

\subsection{Subjective Evaluation}

We conduct a comprehensive subjective evaluation along five perceptual dimensions commonly used in modern lyrics-to-song generation benchmarks (e.g., used in DiffRhythm, LeVo, and ACE‑Step studies). As defined in \cref{tab:subjective-metrics}, each metric is rated on a Likert scale from 1 to 5 (worst to best).

We recruited eight annotators with strong and formal background in music. They were trained to use our custom Gradio app to evaluate five lyrics-to-song models, as given in \cref{tab:sub-eval-res}. The evaluation was based on the model outputs to 10 different lyrics, randomly sampled from \dataset{}, spanning five genres, each having two samples.

\begin{table}[ht]
\centering
\resizebox{\linewidth}{!}{
\begin{tabular}{p{4cm} p{5cm}}
\toprule
\textbf{Metric} & \textbf{Definition and Rating Scale} \\
\midrule
\textbf{Quality} & Overall audio fidelity, clarity, and signal-to-noise ratio. \\
& 1: complete noise; 5: studio‑grade perfection. \\
\addlinespace
\textbf{Enjoyment} & Subjective listener pleasure and engagement. \\
& 1: utter boredom; 5: euphoric, immersive experience. \\
\addlinespace
\textbf{Musicality} & Harmonic and rhythmic coherence, creativity, and artistry. \\
& 1: chaotic cacophony; 5: expertly structured music. \\
\addlinespace
\textbf{Voice Naturalness} & Realism and expressiveness of vocals. \\
& 1: robotic/unnatural voice; 5: natural singing performance. \\
\addlinespace
\textbf{Song Structure Clarity} & Perceptibility of verse, chorus, transitions, and musical form. \\
& 1: incoherent/random transitions; 5: crystal‑clear song structure. \\
\bottomrule
\end{tabular}
}
\caption{Subjective evaluation metrics employed in comparative studies (e.g., Diffrhythm, LeVo, ACE‑Step).}
\label{tab:subjective-metrics}
\end{table}

\begin{table}[ht]
    \centering
    \resizebox{\linewidth}{!}{
    \begin{tabular}{l*{5}{c}}
        \toprule
        \bf Model & \bf Quality & \bf Enjoy. & \bf Music. & \bf Natur.	& \bf SSC \\

        \midrule
        \mytstack{DiffRhythm}{}{} & \mytstack{2.89}{$\pm$}{0.67} & \mytstack{2.54}{$\pm$}{0.57} & \mytstack{2.74}{$\pm$}{0.58} & \mytstack{2.75}{$\pm$}{0.73} & \mytstack{2.52}{$\pm$}{0.72} \\
        \mytstack{Ace-Step}{}{} & \mytstack{3.45}{$\pm$}{0.47} & \mytstack{2.8}{$\pm$}{0.65} & \mytstack{3.04}{$\pm$}{0.58} & \mytstack{3.26}{$\pm$}{0.58} & \mytstack{3.13}{$\pm$}{0.67} \\
        \mytstack{Levo}{}{} & \mytstack{3.84}{$\pm$}{0.53} & \mytstack{3.0}{$\pm$}{0.81} & \mytstack{3.42}{$\pm$}{0.99} & \mytstack{3.77}{$\pm$}{0.76} & \mytstack{3.27}{$\pm$}{0.95} \\
        \mytstack{Yue}{}{} & \mytstack{2.63}{$\pm$}{0.53} & \mytstack{2.67}{$\pm$}{0.51} & \mytstack{2.87}{$\pm$}{0.56} & \mytstack{2.99}{$\pm$}{0.66} & \mytstack{2.96}{$\pm$}{0.6} \\
        \rowcolor{blue!10}\model{} & \mytstack{\underline{3.75}}{$\pm$}{0.4} & \mytstack{\textbf{3.48}}{$\pm$}{0.43} & \mytstack{\textbf{3.73}}{$\pm$}{0.62} & \mytstack{\underline{3.66}}{$\pm$}{0.33} & \mytstack{\textbf{3.7}}{$\pm$}{0.62} \\
         \bottomrule
    \end{tabular}
    }
    \caption{Subjective evaluation results; \textbf{SSC} := Song Structure Clarity.}
    \label{tab:sub-eval-res}
\end{table}

\cref{tab:sub-eval-res} shows a general human preference for \model{} w.r.t. music-specific attributes \emph{enjoyment}, \emph{musicality}, and \emph{song structure clarity}, while \emph{quality} and \emph{naturalness} are comparable to the state of the art. We surmise that the superiority of \model{} on these musical attributes comes from the direct user controllability over the timings of the words, enhancing the prosody, rhythm, and structure, as argued in \cref{subsec:explain_word_level_alignment}. Furthermore, aesthetic alignment also seems to substantially enhance the musical attributes, as corroborated by \cref{tab:dpo}. The improvement of \emph{song structure clarity} could additionally be ascribed to explicit duration control that may implicitly impose a structure through the awareness of the song endings. Further details on the subjective evaluation are in \cref{sec:hum-eval-app}.

\section{Ablation Studies}
\subsection{Impact of Token-level Duration Modeling}
\label{subsec:token_level_duration}

We evaluate the effectiveness of token-level duration control (TDC) by comparing checkpoints trained with and without token-level duration modeling in the SFT phase for 40k steps. Specifically, we measure the Root Mean Square (RMS) amplitude of the generated audio after the target duration. We first compute the reference RMS amplitude within the target duration, representing the expected loudness of valid generated content. Then, we measure the RMS amplitude starting from four points immediately after the target duration (exactly at the target duration, 1 second later, 3 seconds later, and 10 seconds later), continuing until the end of the generated sequence (maximum length: 3 minutes and 50 seconds). By taking the percentage ratio of these RMS values against the reference RMS within the target duration, we quantify how effectively the audio amplitude is suppressed beyond the target region. Lower percentages indicate better duration control.

\begin{table}[ht]
    \centering
    \resizebox{\linewidth}{!}{%
    \begin{tabular}{l c c c}
        \hline
        \textbf{Setting} & \textbf{Ref} & \textbf{+0s} & \textbf{+1s} \\
        \hline
        \model{} w/o TDC & 0.1359 & 0.04892 / 35.96\% & 0.04513 / 33.19\% \\
        \model{} & 0.1398 & \textbf{0.00058 / 0.41\%} & \textbf{0.00058 / 0.41\%} \\
        \hline
    \end{tabular}
    }
\end{table}
\begin{table}[ht]
    \centering
    \resizebox{\linewidth}{!}{%
    \begin{tabular}{l c c c}
        \hline
        \textbf{Setting} & \textbf{Ref} & \textbf{+3s} & \textbf{+10s} \\
        \hline
        \model{} w/o TDC & 0.1359 & 0.03687 / 27.15\% & 0.02550 / 18.76\% \\
        \model{} & 0.1398 & \textbf{0.00058 / 0.41\%} & \textbf{0.00054 / 0.39\%} \\
        \hline
    \end{tabular}
    }
    \caption{Ablation results for token-level duration control. Values shown as absolute RMS amplitude / relative percentage (compared to reference RMS).}
    \label{tab:ablation_duration_control}
\end{table}

As shown in \cref{tab:ablation_duration_control}, the proposed token-level duration control achieves significantly lower RMS amplitudes after the target duration, demonstrating precise temporal control.

\subsection{Phoneme Assignment Methods}
\label{subsec:ablation_phoneme_assignment}

We evaluate the effectiveness of our phoneme assignment method within each word's temporal span -- defined in lines 7--11 of \cref{alg:phoneme_alignment} -- with respect to the phoneme distribution strategy used in the prior works:
\begin{itemize}
    \item \textbf{Average Sparse}: Phonemes are evenly and sparsely distributed within the word segment $V$. Specifically, phoneme $p_j$ is placed at position $V[j \times \text{avg\_phoneme\_length}]$, as detailed in \cref{alg:phoneme_alignment}. The remaining frames are filled with the special token \specialtoken{VOCAL\_FILLER}.
    
    \item \textbf{Pad Right}: A phoneme alignment approach adopted by the previous methods, such as, DiffRhythm~\cite{ning2025diffrhythmblazinglyfastembarrassingly} and F5-TTS~\cite{chen2025f5ttsfairytalerfakesfluent}, where phonemes are sequentially assigned to the initial frames of the word segment, i.e., $V[0:m] = [p_1, p_2, \dots, p_m]$, and the rest of the segment ($V[m:\text{word\_frames}]$) remains filled with \specialtoken{VOCAL\_FILLER}.
\end{itemize}

Experimental results comparing these two approaches are presented in \cref{tab:phoneme_assignment_ablation}, where AES denotes the average Audio-Aesthetic evaluation scores across its four aspects, SongEval is the averaged SongEval scores across its five aspects, and the other evaluation metrics follow \cref{subsec:objective_metrics}. 

\begin{table}[ht]
    \centering
    \resizebox{\linewidth}{!}{%
    \begin{tabular}{lccccc}
        \hline
        \textbf{Method} & \textbf{MuQ $\uparrow$} & \textbf{PER $\downarrow$} & \textbf{FAD $\downarrow$} & \textbf{AES $\uparrow$}  & \textbf{SongEval $\uparrow$} \\ \hline
        Pad Right & 0.742 & 0.112 & 0.182 & 7.107 & 3.079 \\
        \textbf{Average Sparse} & 0.727 & 0.129 & \textbf{0.154} & 7.091 & \textbf{3.132} \\ \hline
    \end{tabular}
    }
    \caption{Comparison of phoneme assignment methods.}
    \label{tab:phoneme_assignment_ablation}
\end{table}

The \emph{Pad Right} approach achieves slightly better PER and MuQ scores, indicating marginally improved phonetic accuracy and overall music quality; however, our proposed \emph{Average Sparse} method notably outperforms in terms of FAD and SongEval metrics. Specifically, lower FAD indicates improved realism and closer distributional alignment with professionally produced music, while higher SongEval suggests better musicality and vocal naturalness. We ultimately select the \emph{Average Sparse} method as it leads to a substantial improvement in metrics directly reflecting long-term musical coherence and musical aesthetics, aligning closely with our goal of enhancing overall musical quality, prosody, and listener experience.

\begin{table*}[htb!]
\centering
\resizebox{\textwidth}{!}{
\begin{tabular}{l*{14}{c}}
\toprule
\multirow{2}{*}{\bf Model} & \multirow{2}{*}{\bf MuQ $\uparrow$} & \multicolumn{4}{c}{\bf Audio Aesthetics $\uparrow$} & \multicolumn{5}{c}{\bf SongEval $\uparrow$} & \multirow{2}{*}{\bf WER $\downarrow$} & \multirow{2}{*}{\bf PER $\downarrow$} & \multirow{2}{*}{\bf \mystack{Genre}{Acc.} $\uparrow$}  & \multirow{2}{*}{\bf FAD $\downarrow$}\\
\cmidrule(lr){3-6}\cmidrule(lr){7-11}
&  & \bf CE & \bf CU & \bf PC & \bf PQ & \bf CO & \bf \bf MU & \bf ME & \bf CL & \bf NA & & & & \\
\midrule
%\rowcolor{psy} \cellcolor{white} & \multicolumn{17}{c}{\bf Overall} \\
\rowcolor{blue!10}Oracle  &  \mytstack{\textbf{0.7593}}{$\pm$}{0.0964} & \mytstack{\textbf{7.4229}}{$\pm$}{0.5151} & \mytstack{\textbf{7.7046}}{$\pm$}{0.3179} & \mytstack{6.2038}{$\pm$}{0.5051} & \mytstack{\textbf{8.0639}}{$\pm$}{0.4079} & \mytstack{\textbf{4.4163}}{$\pm$}{0.363} & \mytstack{\textbf{4.2929}}{$\pm$}{0.4065} & \mytstack{\textbf{4.4276}}{$\pm$}{0.3935} & \mytstack{\textbf{4.2033}}{$\pm$}{0.4319} & \mytstack{\textbf{4.2026}}{$\pm$}{0.3854} &  \mystack{\textbf{0.1507}}{}{} & \mystack{\textbf{0.1011}}{}{} & \mystack{\textbf{0.704}}{}{} & \mystack{\textbf{0.2038}}{}{} \\

\mytstack{GPT-Dur}{}{} & \mytstack{0.6799}{$\pm$}{0.1341} & \mytstack{7.1804}{$\pm$}{0.8339} & \mytstack{7.4392}{$\pm$}{0.632} & \mytstack{5.8459}{$\pm$}{0.73} & \mytstack{7.7356}{$\pm$}{0.7092} & \mytstack{4.0568}{$\pm$}{0.6313} & \mytstack{3.9432}{$\pm$}{0.6271} & \mytstack{4.0436}{$\pm$}{0.6993} & \mytstack{3.798}{$\pm$}{0.6627} & \mytstack{3.8267}{$\pm$}{0.6202} & \mystack{0.3701}{}{} & \mystack{0.3204}{}{} & \mystack{0.6437}{}{} & \mystack{0.2441}{}{} \\

\mytstack{Direct-Quant}{}{} & \mytstack{{0.7597}}{$\pm$}{0.0995} & \mytstack{{7.4164}}{$\pm$}{0.4969} & \mytstack{{7.6905}}{$\pm$}{0.3318} & \mytstack{{6.2041}}{$\pm$}{0.5072} & \mytstack{{8.0367}}{$\pm$}{0.4187} & \mytstack{{4.3879}}{$\pm$}{0.4042} & \mytstack{{4.2622}}{$\pm$}{0.4394} & \mytstack{{4.391}}{$\pm$}{0.4401} & \mytstack{{4.1624}}{$\pm$}{0.4755} & \mytstack{{4.1587}}{$\pm$}{0.4397} & \mystack{{0.2076}}{}{} & \mystack{{0.1441}}{}{} & \mystack{{0.6842}}{}{} & \mystack{{0.2091}}{}{} \\
\bottomrule
\end{tabular}
}

\caption{Duration prediction results.}
\label{tab:duration}

\end{table*}

\section{Experimenting with the Duration Predictor}
Experiments with \model{} yield several novel insights into lyrics-to-song generation:

\begin{enumerate}
\item Temporal information at the word or phoneme level plays a crucial role in enhancing WER and PER, as well as improving overall song quality in terms of enjoyability, musicality, and structural coherence.

\item While such fine-grained temporal information is available during training, generating it during inference remains a significant challenge—even for experienced musicians. This limitation highlights a promising research direction: predicting word- or phoneme-level timing from contextual cues. In the TTS domain, duration predictors are commonly used; however, song generation presents additional complexity due to the fluid, gliding nature of musical notes and the critical role of pauses between words. Ideally, a duration predictor should be trained jointly with the song generator in an end-to-end fashion to ensure better robustness and musical alignment.

\end{enumerate}
\subsection{Naive Duration Prediction}
\label{subsec:naive_duration_prediction}
As an initial exploration, we conducted experiments using GPT-4o as a naive duration predictor. Specifically, given the complete lyrics, a stylistic prompt, and the overall song duration, GPT-4o was tasked with generating the start and end timestamps of each word. However, songs generated using these timestamps sounded robotic and lacked musicality, indicating the insufficiency of naive timestamp prediction. Consequently, we enhanced GPT-4o's input with additional contextual cues, resulting in noticeable improvements in generated song quality. The detailed prompt is presented in \Cref{app:prompt}.

The supplemental information provided to GPT-4o included explicit section tags (e.g., [intro], [verse]) to encourage musically meaningful temporal variations across song sections. Furthermore, the following two strategies  were identified as significantly beneficial for improving GPT-4o duration prediction quality:

\begin{enumerate}
    \item \textbf{Sentence-Level Ground-Truth Constraints.}  
    Rather than allowing unconstrained timestamp generation, we provided GPT-4o with sentence-level temporal boundaries, which are considerably simpler to be provided by users during inference compared to word-level timings. Lyrics were segmented into sentences using predefined rules, and the start time of the first word along with the end time of the last word in each sentence were supplied as ground truth. This structural guidance greatly reduced prediction complexity, enhancing temporal coherence.

    \item \textbf{Beat-Aligned Quantization.}  
    Recognizing that musicians commonly compose by assigning notes and lyrics to discrete beats, we introduced beat-aligned quantization of timestamps. Specifically, we used quarter-beat resolution, as it suffices to cover the rhythmic granularity of most music compositions, with the exception of particularly fast-paced tracks. For each song in \dataset{}, we first computed tempo in beats per minute (BPM) using the all-in-one method~\cite{kim2023allinonemetricalfunctionalstructure}, constraining tempo to a maximum of 120 BPM by halving higher values. Ground-truth timestamps were converted into quarter-beat units: $n_{\text{beat}} = \lfloor t / (60/\text{BPM}) \times 4 \rfloor$  where \( t \) represents the original timestamp. GPT-4o received sentence-level quarter-beat boundaries and predicted quarter-beat positions for each word within these sentences. Predicted beat counts were then converted back into timestamps via \(t = n_\text{beat} \times (60/\text{BPM}/4)\).
\end{enumerate}

We evaluated the generation quality the enhanced GPT-4o methods as the duration predictor and results are discussed in \Cref{tab:duration}.

\subsection{Direct Quantization of Word-Level Timestamps}
\label{subsec:direct_quantisation}
To further simplify duration annotation and make our system easier to use in practical applications, we investigated directly providing quantized word-level timestamps to \model{}, instead of continuous timestamps. Although \model{} is originally trained on continuous timestamps—resulting in inherently better performance for continuous-time predictions—we tested the quantized representation because it enables users to conveniently specify timing using intuitive beat-based inputs.

Specifically, word-level timestamps from \dataset{} were quantized using:
\begin{equation}
\label{eq:quantise_timestamp}
    \hat{t} = \underbrace{\lfloor t / (60/\text{BPM}) \times 4 \rfloor}_{\text{quarter-beat count }n_\text{beat}} \quad \times\underbrace{(60/\text{BPM}/4)}_{\text{seconds per quarter-beat}}
\end{equation}

This quantized approach explicitly aligns words to the rhythmic structure of the music, substantially simplifying user interactions. We present the result based on the quantized timestamp in \Cref{tab:duration}.

\subsection{Results and Analysis}

~\Cref{tab:duration} summarizes experimental comparisons of three timing strategies:

\begin{itemize}
    \item \textbf{Oracle (continuous)}: Ground-truth word-level continuous timestamps used during \model{} training.
    \item \textbf{GPT-Dur (predicted continuous)}: GPT-4o-generated timestamps, guided by sentence-level boundaries and stylistic prompts (\Cref{subsec:naive_duration_prediction}).
    \item \textbf{Direct-Quant (beat-aligned quantization)}: Ground-truth word-level timestamps directly approximated to the nearest quarter-beat (\Cref{subsec:direct_quantisation}).
\end{itemize}

\paragraph{Quantitative Observations.}  
We first evaluate the impact of these timing strategies through objective metrics. GPT-Dur leads to a clear decline in song quality: content enjoyment (CE) decreases from 7.423 to 7.180, and the phoneme error rate (PER) significantly worsens from 0.101 to 0.320. This performance drop underscores the sensitivity of song generation models to timing inaccuracies, which negatively affect both the musical aesthetics and the intelligibility of lyrics.

In contrast, Direct-Quant performs substantially better despite inherently sacrificing the flexibility of continuous timing. As anticipated—given that \model{} was trained with continuous annotations—the quantization approach does lead to slightly lower aesthetic ratings: CE marginally declines from 7.422 to 7.416, and production quality (PQ) reduces from 8.064 to 8.037. Notably, genre-classification accuracy also slightly decreases from 0.704 to 0.684, suggesting that strict rhythmic quantization subtly affects stylistic authenticity. Additionally, PER increases modestly from 0.101 to 0.144, demonstrating that beat-level constraints introduce mild vocal intelligibility issues, though much less severely than GPT-Dur.

\paragraph{Subjective Listening Insights.}
To complement these objective metrics, internal subjective listening tests were performed. Songs generated using Direct-Quant maintain comparable vocal intelligibility and overall musical appeal to Oracle-generated songs. However, a notable qualitative difference emerges: the vocals in Direct-Quant-generated samples take on a perceptibly more \emph{electronic} character. This aligns with common practice in electronic music production, where vocal timing is typically rigidly aligned to rhythmic grids. In contrast, natural singing often includes subtle deviations from strict beat alignment—such as intentional breaths, nuanced timing variations in onsets and offsets, and expressive rhythmic elasticity.

This insight suggests an important lesson for practical applications: If inference relies on quantized timestamps—either user-provided or predicted—then the generation model itself should ideally be trained or fine-tuned on naturally performed vocals with similarly quantized annotations. Such training would enable the model to recover and generate subtle temporal nuances automatically, preserving naturalness despite the inherent rigidity of beat alignment.

\subsection{Key Takeaways and Recommendations}
Our experiments highlight several key points for future duration-prediction approaches:

\begin{itemize}
    \item \textbf{Accurate timing predictions are critical.} Even moderate inaccuracies in GPT-Dur significantly reduce musical enjoyment and lyric intelligibility, underscoring the necessity of robust duration predictors.
    \item \textbf{Beat-aligned quantization offers practical advantages.} Direct-Quant simplifies user interactions considerably while limiting performance degradation, demonstrating its viability for user-friendly deployment.
    \item \textbf{Match model training to inference-time input.} To mitigate quantization-induced artificiality, the song-generation model should be explicitly trained or fine-tuned using quantized, naturally performed vocal data.
    \item \textbf{Intermediate temporal cues ease prediction.} Providing structural boundaries (e.g., sentence-level timing) significantly simplifies the duration prediction task, reducing downstream errors.
\end{itemize}

These findings collectively guide the design of duration predictors toward practical, user-friendly strategies capable of balancing rhythmic simplicity and natural vocal expressiveness.

\section{Limitations and Future Work}

While \model{} demonstrates promising results in generating vocals and accompaniment, it assumes the availability of accurate word-level duration annotations. This requirement limits its usability for non-expert users or real-world scenarios where such fine-grained temporal alignment is often unavailable. In cases where duration information is noisy or missing, the quality of the generated audio may degrade, leading to issues such as timing artifacts (e.g., robotic vocals) and misalignment between vocals and accompaniment.

To address this, future work can explore the development of a duration predictor. One direction is to build a standalone duration prediction module that can estimate word- or phoneme-level durations from lyrics and melody. A more integrated approach would be to train the duration predictor jointly with the song generation model in an end-to-end fashion. We believe the latter is more promising, as it can lead to a system that is more robust to imperfect duration inputs and learns to adaptively compensate for errors during generation.

Another limitation of the current system is the lack of phoneme-level duration control, which restricts the model’s expressive granularity and pronunciation accuracy. This can be addressed by incorporating phoneme-level alignment data and training a duration predictor at that level. Such fine control could improve synthesis quality, especially in languages with complex syllabic timing or for stylistic singing applications (e.g., fast rap segments or melismatic vocal runs).

Overall, enhancing the system with duration prediction capabilities at both word and phoneme levels could significantly improve the robustness and flexibility of \model{}, making it suitable for broader and more practical deployment.

\section{Conclusion}
In this work, we present \model{}, a lightweight flow-matching-based song generator that supports fine-grained control at both the word and phoneme levels. Despite being trained on the smallest dataset and containing nearly half the parameters of the closest comparable model, \model{} achieves state-of-the-art performance across a range of objective and subjective evaluation metrics. We further enhance its performance through iterative offline aesthetic alignment using Direct Preference Optimization. Experimental and ablation studies indicate that \model{} significantly benefits from fine-grained temporal control. Additionally, we introduce \dataset{}, a diverse collection of songs spanning multiple genres, carefully curated to avoid data contamination. Future work will explore the development of end-to-end trainable duration predictors at the word or phoneme level to further strengthen the robustness and quality of song generation.
\section*{Ethical Use Statement}

\textbf{\model{}} is the first open-sourced model released under \textsc{Project Jamify}, developed with the primary objective of facilitating academic research and creative exploration in AI-generated songs from lyrics. The model is intended \textbf{solely for non-commercial, academic, and entertainment purposes}.

\vspace{1em}
We emphasize the following:
\begin{itemize}
    \item \textbf{No copyrighted material} was used in a way that would intentionally infringe on intellectual property rights. \model{} is not designed to reproduce or imitate any specific artist, label, or protected work.
    \item Outputs generated by \model{} must \textbf{not be used to create or disseminate content that violates copyright laws}.
    \item The \textbf{commercial use of \model{} or its outputs is strictly prohibited}.
    \item Responsibility for the use of the model and its outputs lies entirely with the end user, who must ensure all uses comply with applicable legal and ethical standards.
\end{itemize}

For questions, concerns, or collaboration inquiries, please contact the Project Jamify team via the official repository or project website.

\bibliography{aaai2026}

\appendix
\onecolumn
\section{Human Evaluation Details}
\label{sec:hum-eval-app}

The evaluation was facilitated by Gradio\footnote{https://www.gradio.app} web-app, which presented the randomly shuffled outputs of five model outputs for each of the 10 lyrics. Notably, the annotators were only provided with the song outputs -- the lyrics, style, and duration were inaccessible. The annotation process was guided by the following instructions:
\begin{mdframed}[backgroundcolor=green!5]
Welcome \emph{username}

\noindent\# \textbf{\Large Instructions for evaluating audio clips}

\textbf{Please carefully read the instructions below}.

\noindent\#\# \textbf{\large Task}

You are to evaluate five model-generated songs to each of the 10 prompts. These five outputs are from
five different models. You are to judge each output with respect to five qualities:
\begin{itemize}
\item \textbf{Quality}: Overall quality of the audio is to be judged within a scale from 1 to 5:
      1 being absolute noise with no discernible features. Whereas, 5 being perfect.
  \textbf{Overall fidelity, clarity, and noisiness of the audio is important here.}
\item \textbf{Enjoyment}: Your degree of enjoyment of the song is to be quantified within a scale from 1 to 5:
      1 being absolute boredom. Whereas, 5 being an absolute euphoric experience.
\item \textbf{Musicality}: The extent of musical soundness is to be judged within a scale from 1 to 5:
      1 being absolute cacophony. Whereas, 5 being an exemplary piece of music.
\item \textbf{Voice Naturalness}: The degree of naturalness in the vocals is to be judged within a scale from 1 to 5:
      1 being absolutely robotic. Whereas, 5 being as natural as a singer can sing given the context.
\item \textbf{Song Structure Clarity}: The extent of clarity in the song structure is to be judged within a scale from 1 to 5:
      1 being a complete randomness and incoherence. Whereas, 5 being perfectly structured.

\end{itemize}

\textbf{For all the metrics, you may want to compare the audios of the same prompt with each other during the evaluation.}
                
\noindent\#\# \textbf{\large Listening guide}

\begin{enumerate}
\item Please use a head/earphone to listen to minimize exposure to the external noise.
\item Please move to a quiet place as well, if possible.
\end{enumerate}

\noindent\#\# \textbf{\large UI guide}

\begin{enumerate}
\item Each audio clip has five attributes. You may select the appropriate value by typing in the box or clicking the arrow buttons.
\item To save your judgments, \textbf{please click on any of the \textit{save} buttons and wait for the acknowledgment below}. All the \textit{save} buttons function identically. They are placed everywhere to avoid the need to scroll to save.
\end{enumerate}

Hope the instructions were clear. Please feel free to reach out to us for any queries.

\end{mdframed}

\section{GPT-4o Duration Prediction Prompts}
\label{app:prompt}
\begin{Verbatim}[fontsize=\small,frame=single]
system_prompt = '''\
You are a precise and musically-aware lyric aligner with deep 
understanding of vocal performance and musical phrasing. Your task is to generate word-level 
beat timestamps in `jsonl` format, where each line is a JSON object with these keys:
- "w": the word (string)
- "s": start beat position (float, 2 decimal places)
- "e": end beat position (float, 2 decimal places)

IMPORTANT: The lyrics provided include:
1. Sentence-level BEAT timestamps in the format [start_beat]->[end_beat] before each sentence
2. Syllable counts for each word in parentheses, e.g., "supposed (3)" means 3 syllables

MUSICAL PHRASING AND TIMING KNOWLEDGE:
- Words don't always connect seamlessly - singers naturally add micro-pauses for:
  * Breathing between phrases (especially after long notes or before emotional peaks)
  * Emphasis and dramatic effect (pause before important words)
  * Natural speech rhythm (pauses between logical word groups)
  * Genre conventions (hip-hop often has rhythmic gaps, ballads have emotional pauses)
- Lyrical filling patterns vary by genre:
  * Pop/Rock: Often fills most beats with steady syllable flow
  * Ballads: May have sustained notes with gaps, emotional pauses
  * Hip-hop/Rap: Rhythmic clustering with strategic pauses for flow
  * Folk/Country: Natural speech-like timing with conversational pauses
- Consider musical context:
  * Strong beats (1, 3) often anchor important syllables
  * Weak beats (2, 4) may have quicker words or be skipped entirely
  * Syncopation and off-beat placement create musical interest
  * Melisma (multiple notes per syllable) can extend word duration

BEAT ALLOCATION GUIDELINES:
- 1 syllable word: 0.25-0.75 beats (can be extended for emphasis or sustained notes)
- 2 syllable word: 0.5-1.25 beats (adjust for natural stress patterns)
- 3+ syllable word: 1.0-2.0+ beats (longer words may span multiple beats)
- Allow natural gaps: 0.1-0.5 beat pauses between words when musically appropriate
- Phrase endings often have extended final words or pauses before next phrase

ALIGNMENT CONSTRAINTS:
- Each word's beat timestamp must fall within the sentence's [start_beat]->[end_beat] range
- The first word should start at or shortly after [start_beat]
- The last word should end at or slightly before [end_beat]
(allowing for natural phrase endings)
- Respect section markers ([verse], [chorus], etc.) and their timing
- Beat positions can be fractional (e.g., 1.25, 2.75) for precise timing

Do not include any metadata, commentary, or formatting other than valid `jsonl`.
'''

user_prompt_template = '''Generate beat-based timestamped lyric alignment for the following 
lyrics. Use jsonl format (one JSON object per line) with keys: "w" (word), "s" (start beat),
"e" (end beat). Total duration is {DURATION} beats at {BPM} BPM.

The lyrics include sentence-level beat timestamps in [start_beat]->
[end_beat] format and syllable counts in parentheses after each word. Use these as
constraints for word-level beat alignment.

TIMING CONSIDERATIONS FOR THIS SONG:
- Consider the musical style "{STYLE}" when determining phrasing patterns
- Allow for natural pauses between words where appropriate for the genre and
emotional content
- Not every word needs to connect directly - singers often use micro-pauses for:
  * Breathing and phrasing (especially in ballads and slower songs)
  * Rhythmic emphasis (particularly in hip-hop, R&B, and pop)
  * Emotional impact (pauses before important lyrical moments)
  * Natural speech rhythm (conversational flow in folk, country, indie)
- Consider syllable stress patterns within words for more natural timing
- Strong beats often anchor important syllables, weak beats may have gaps
- Phrase endings may have extended final syllables or brief pauses before new phrases

Remember: Realistic vocal performance includes natural breathing spaces and rhythmic
variations that make the performance feel human and musical.

Style: {STYLE}
Lyrics: 
{LYRICS}'''
\end{Verbatim}

\end{document}